# Evidence of spin-phonon-charge coupling in quasi-one-dimensional Ising spin chain system α – CoV$_2$O$_6$

Debismita Naik,[1,*] Souvick Chakraborty,[1] Akriti Singh,[2] Ayan Mondal,[3] Surajit Saha,[2] Venkataramanan Mahalingam,[3] Satyabrata Raj,[1,4] and Pradip Khatua[1,†]

[1]Department of Physical Sciences, Indian Institute of Science Education and Research Kolkata, Mohanpur, Nadia, West Bengal 741246, India.

[2]Department of Physics, Indian Institute of Science Education and Research, Bhopal 462066, India.

[3]Department of Chemical Sciences, Indian Institute of Science Education and Research Kolkata, Mohanpur, Nadia, West Bengal 741246, India.

[4]National Centre for High-Pressure Studies, Indian Institute of Science Education and Research Kolkata, Mohanpur, Nadia, 741246, India

Corresponding authors: [*]dn17ip012@iiserkol.ac.in

[†]pradip.k@iiserkol.ac.in

## ABSTRACT

The quasi-one-dimensional Ising spin chain system α-CoV$_2$O$_6$ is considered to exhibit fascinating magnetic properties at lower temperatures. We comprehensively study magnetic properties and lattice dynamics using a combination of x-ray diffraction (XRD), DC magnetization, specific heat, temperature-dependent XRD and Raman scattering measurements, along with the theoretical charge density map calculations. The DC magnetization and specific heat confirm the antiferromagnetic (AFM) long-range order (LRO) at Neel temperature $T_N$ = 15 K as well as the short-range ordering (SRO) below 100 K. Temperature-dependent Raman measurement has been performed in the temperature range of 5 K to 180 K, revealing spin-phonon coupling (SPC) below 100 K, which is well above the $T_N$ and is attributed to SRO. The SPC is calculated for different vibrational modes. The temperature-dependent XRD supports the key finding of magnetoelastic coupling at $T_N$, which is strongly correlated to the magnetoelectric phenomena. The renormalization of Raman modes and lattice anomalies near $T_N$ illustrate spin-lattice coupling via magnetoelastic and spin-phonon interactions leading to the interplay between spin, charge, and phonon degrees of freedom in α-$CoV_2O_6$. The presence of spin-charge coupling is confirmed by the theoretical charge density maps, which show that the formation of electric dipoles between the Co and O atoms in the AFM state results from the p-d hybridization.

# INTRODUCTION

Low-dimensional magnetic spin systems have attracted high interest due to their complex interplay between spin, orbit, charge and lattice degrees of freedom, such as a type of system is crucial for achieving a range of functional properties, including magnetoelectric effect, ferroelectricity, superconductivity, metal-insulator transition, and charge density wave. In these materials, the magnetic ions are arranged in linear chains, planes, or ladders [1–3]. The low-dimensional magnetic sublattice can result in a complex phase diagram, where minor variations in temperature and magnetic field trigger a rearrangement of the magnetic order. One-dimensional magnetic oxide α-$CoV_2O_6$ exhibits fascinating phenomena such as large orbital moment, 1/3rd magnetization plateau, giant magnetostriction, and spin-orbital transition [4–7]. Geometrical frustration due to the triangular arrangement of magnetic moments results in competition between the antiferromagnetic (AFM) and ferromagnetic (FM) exchange interaction [8–10]. These properties are the basis of potential technological applications, and these materials exhibit a magnetic memory effect on the nanoscale, which is used for digital information storage application purposes [11]. Multiple studies have shown that the field-induced metamagnetic transition shows a significant magnetic entropy change and magnetocaloric effect (MCE), which is a notable advancement in refrigeration technology [1].

$CoV_2O_6$ crystallizes in two phases depending on the local environment of the $Co^{2+}$ ion. One is a high-temperature monoclinic α-phase and another is low-temperature triclinic γ-phase with C2/m and P-1 space group respectively [8,12]. In α-$CoV_2O_6$, high spin $Co^{2+}$ in edge-sharing chains of highly distorted $CoO_6$ octahedra forms a linear chain running along the b-axis and separated by a non-magnetic vanadium oxide layer composed of zigzag chains of $VO_6$ octahedra but by $VO_6$ octahedra and $VO_4$ tetrahedra in the triclinic phase [2,13]. Although the spin chains are similar in nature, the direction of their magnetization easy axis is different. The magnetization easy axis is along the c axis in the α phase whereas along the chain direction in the γ phase. As the $V^{5+}$ ion is a nonmagnetic state, the magnetic contribution comes from the $Co^{2+}$ ion. Both α and γ - $CoV_2O_6$ phase exhibit long range antiferromagnetic ordering below 15 K and 6 K respectively and show field-induced transitions. One-third of saturation magnetizations step occurs from AFM to FM state via Ferrimagnetic (FIM) state at two critical fields when the magnetic field is applied along the c-axis. The magnetic moments of $Co^{2+}$ ions along the b-axis from an FM chain and interchain are antiferromagnetically coupled in an ac-plane [12]. The Co-Co distance in the FM chain is small compared to the AFM chain, so the FM chain is more strongly coupled than the AFM chain. The $Co^{2+}$ chains are arranged in an isosceles triangular geometry. Still, in an ac-plane, these chains are arranged at different lengths, which leads to frustration showing a plateau in field-induced magnetization. The ferromagnetic moment of 4.5 $\mu_B$/ $Co^{2+}$ is larger than the spin-only value in α-$CoV_2O_6$ indicating a large value of orbital contribution due to the high spin state of $Co^{2+}$ (S = 3/2) [14]. The Neutron triple-axis spectroscopy (TASP) validated the Ising nature of α-$CoV_2O_6$ from the powder diffraction analysis [15]. This large orbital moment of quasi-1D Ising spin system α-$CoV_2O_6$ arises from the cooperative effects of the spin-orbit coupling (SOC) and crystal electric field of highly distorted $CoO_6$ octahedron [16]. The DFT calculations have correctly predicted the AFM ground state and FM exchange value of 30 K along the chain and give an acceptable description of the magnetic insulator. The theoretical calculation using GGA + SOC approximation could not evaluate the orbital moment [17]. So, the local magnetism was reported using x-ray magnetic circular dichroism (XMCD) spectroscopy to get separate information on spin and orbital moments [12]. From the Neutron diffraction study, it was found a large magnetostriction and volume expansion with both magnetic field and temperature suggesting strong spin-lattice coupling in α-$CoV_2O_6$ [4,7]. Dielectric measurements revealed the evidence of coupling between magnetic ordering and electrical charges [3]. A recent study suggested the spin-lattice coupling plays a crucial role in maintaining an equilibrium plateau [18]. Neutron diffraction, dielectric, magnetostriction, and magnetic circular dichroism experiments have shown a strong coupling between spin, charge, lattice, and orbital in α-$CoV_2O_6$.

These observations further motivated us to explore the detailed study of spin-phonon coupling which has remained undiscovered in this spin-chain material. For this purpose, we performed Raman scattering measurements to study the lattice vibrations and the presence of SPC. The SPC can also serve as a valuable tool for investigating various intriguing phenomena, including the magnetoelectric effect, magnetostriction, etc. The SPC is a compelling phenomenon that arises from the renormalization of phonon modes due to magnetic ordering. Our study aims at deeper insight into the physical properties of α-$CoV_2O_6$, particularly its detailed magnetic, structural, and lattice dynamics, thereby expanding beyond the current knowledge of its magnetization, specific heat results, x-ray diffraction and Raman spectroscopy studies. The temperature-dependent XRD results further prompt us to calculate charge transfer using DFT calculation (p-d hybridization) for validating the magnetoelectric properties, in which spin-charge-lattice coupling tends to dielectric anomaly at $T_N$.

This article provides an in-depth analysis of α-$CoV_2O_6$ using structural, magnetic, and temperature-dependent Raman scattering. Temperature-dependent magnetization measurement and specific heat reveal LRO at $T_N$ = 15 K and SRO below 100 K. We have observed six Raman-active modes deviate from the expected cubic-anharmonic trend below 100 K, which can be linked to the SPC. The strength of SPC is estimated by mean-field theory and two-spin cluster method. Our theoretical calculation indicates that strong *p-d* hybridization facilitates charge transfer between the Co-d and O-p orbitals leading to spin-phonon-charge coupling in this system.

## EXPERIMENTAL DETAILS

Polycrystalline samples of α-$CoV_2O_6$ were synthesized by using the sol-gel chemical method [19]. Room temperature powder X-ray diffraction (PXRD) measurement was done by Rigaku x-ray diffractometer with Cu-$K_α$ radiation of wavelength 1.5406 Å. Stoichiometric compositions were determined using the energy dispersive X-ray (EDAX) technique coupled with a high-resolution field emission scanning electron microscope (HR-FESEM). Low-temperature powder X-ray diffraction (LT-PXRD) was done using Smart Lab, RIGAKU Diffractometer with Helium cryostat, which uses a two-stage closed cycle cooler, the temperature goes to 10 K and Cu Kα radiation source with wavelength 1.5418 Å was used. The structural parameters were refined using FULLPROF software using a linear background. Temperature variation DC magnetization measurement with magnetic field was done using Quantum Design SQUID (Superconducting Quantum Interference Device). Heat capacity measurement was performed using a Physical Properties Measurement System (PPMS, Quantum Design) on thin and flat pellet samples in the temperature range of 2 – 300 K. First the data were collected for Apiezon grease which is used to stick the sample to the holder as an addenda in the temperature range of 2 – 300 K during the heating cycle. Measurement was done by mounting the sample (stick to the addenda) in the same temperature range following a thermal relaxation method. The final heat capacity ($C_p$) of the sample was obtained after subtracting the addenda heat capacity from the total heat capacity data. Raman measurement was performed using an HR Evolution Lab RAM Raman spectrometer with laser excitations of 532 nm in the backscattering configuration and a closed cycle attoDRY 1000 cryostat was employed to perform Raman measurements down to 5 K.

## COMPUTATIONAL DETAILS

First-principles Density functional theory (DFT) calculations with and without considering the SOC effect were performed with the help of the Vienna Ab Initio Simulation Package (VASP) software [20] package using the Perdew-Burke-Ernzerhof (PBE) exchange-correlation functional [21]and the frozen-core Projector Augmented Wave (PAW) formalism [22]. The energy cutoff for the plane wave basis was set at 520 eV for all the calculations. The lattice parameters and atomic coordinates have been taken from experimental data. For the antiferromagnetic state, the experimental lattice parameters at 4 K were chosen [7], and for the non-magnetic calculations, we considered the room temperature lattice constants. A *k*-point grid of 6×14×8 was used for all the self-consistent calculations with the energy convergence tolerance set at $10^{-6}$ eV. The GGA+U method with $U_{eff}$ = 5 eV for the Co *d*-atoms was used for all calculations to match the experimental band gap of ~2 eV. The Bader charge analysis was performed with the help of the Bader Charge Analysis code [23]. The crystal structures, charge densities, and charge density differences are plotted by VESTA software [24].

## RESULTS AND DISCUSSIONS

### A. Experimental results

The Rietveld refinement of the powder XRD pattern of α-CoV$_2$O$_6$ confirms a monoclinic space group $C2/m$ at room temperature with lattice parameters a = 9.2543 Å, b = 3.50454 Å and c = 6.6213 Å and α = γ = 90°, β = 111.61° and refined parameter $\chi^2$ = 1.08 manifest a good-quality fitting of the pure phase XRD pattern. For visualization of the crystal structure of α-CoV$_2$O$_6$, VESTA software. All structural parameters using atomic positions and bond distances and all the details of material characterization including elemental composition, as given in Supplemental

Materials [25], are found to be in good agreement with previously reported data. Figure 1(b). illustrates a schematic representation of the crystal structure. The structure consists of $CoO_6$ octahedra forming an intra-chain in the b-direction and inter-chain along the a- and c-direction.

Now we will shift our focus to examine the magnetic properties of the $\alpha$-$CoV_2O_6$. The temperature dependence of magnetization (M) measurement following the standard zero field cooled (ZFC) and field cooled (FC) protocols were measured under a magnetic field (H) of 1 T, as shown in Figure. 2(a). Magnetic susceptibility ($\chi$) shows a long-range antiferromagnetic ordering at $T_N$ = 15 K [shown in Figure. 2(a)]. The inverse susceptibility ($\chi^{-1}$) vs T curve is shown in Figure. 2(b) deviates below 100 K, which was fitted with the modified Curie-Weiss (MCW) law. This behavior implies that there exists a short-range ordering present well above $T_N$.

$$\chi = \chi_0 + \frac{C}{T - \theta_{CW}} \qquad (1)$$

Here $\theta_{CW}$ is the Curie temperature, C refers to the Curie constant and $\chi_0$ represents the temperature-independent contributions including the core-electron diamagnetism and open-shell Van Vleck (VV) paramagnetism. The obtained negative value of $\theta_{CW}$ ~ - 42 K signifies the AFM interaction and $\chi_0$ ~ $10^{-4}$ (order in emu/mole Oe). The calculated effective paramagnetic (PM) moment ($\mu_{eff}$ = 5.7$\mu_B$) through the equation of MCW law for the octahedrally distorted $Co^{2+}$ is larger than the spin-only magnetic contribution value, which indicates the presence of orbital moment leads to SOC [1]. Also, the higher value of the experimental $\mu_{eff}$ is due to SRO being well above the $T_N$ [26]. The existence of SRO is also clarified by the specific heat data.

The specific heat measurement was conducted to explore the nature of magnetic ordering in $\alpha$-$CoV_2O_6$. The $C_P$ data collected at zero field is shown in Figure. 3(a), revealing a sharp distinct anomaly at $T_N$ ~ 15 K, which confirms the onset of long-range order AFM behavior below 15 K. This finding is consistent with the bulk magnetization study and the microscopic spin structure analysis performed through neutron powder diffraction (NPD). In insulating magnetic systems, the $C_P$ consists of two main components: the phononic contribution at high temperatures and the magnetic contribution at lower temperatures. To determine the magnetic specific heat ($C_{mag}$), we subtracted the phononic contribution ($C_{lattice}$) from the total $C_P$. We estimated the phonon contribution by fitting the high-temperature region of the specific heat using a combination of Debye and Einstein model, as expressed below [27]:

$$C_{lattice} = C_{Debye}(T) + C_{Einstein}(T)$$

$$= n_D\left[ 9R\left(\frac{T}{\theta_D}\right)^2 \int_0^{\frac{\theta_D}{T}} \frac{x^4 e^x}{(e^x-1)^2} + \sum_i 3Rn_{E_i}\left(\frac{\theta_{E_i}}{T}\right)^2 \frac{e^{\frac{\theta_{E_i}}{T}}}{\left(e^{\frac{\theta_{E_i}}{T}}-1\right)^2} \right] \quad (2)$$

Where $R$, $\theta_D$, $\theta_{E_i}$ are gas constant, Debye temperature and Einstein temperatures respectively. The first term Debye model represents acoustic phonons while the second term Einstein model corresponds to optical phonons. The factors $n_D$ and $n_{E_i}$ are the respective weighting coefficients. In this context, $n_D + \sum_i n_{E_i} = n = 9$, which corresponds to the total number of atoms per formula unit of α-CoV$_2$O$_6$. Using the combined model, In the high-temperature region 80 - 300 K, the resultant fitting parameters are obtained as Debye term ($n_D = 1$ $\theta_D = 450\ K$) and Einstein terms ($n_{E_1} = 4$, $n_{E_2} = 3$, $n_{E_3} = 1$ and $\theta_{E_1} = 264\ K$, $\theta_{E_2} = 833\ K$, $\theta_{E_3} = 300\ K$). The temperature dependence of the $C_{mag}$ curve, after subtracting the lattice contribution, is shown in Figure. 3(b). This curve $C_{mag}/T$ was further integrated to calculate the magnetic entropy ($S_{mag}$), as shown in Figure. 3(b). However, only about 69% of the total $S_{mag}$ was released at $T_N$, with the remainder being recovered above $T_N$. Additionally, $S_{mag}$ reached a saturation value of approximately 7.24 J/mol·K, which is lower than the expected total magnetic entropy for Co$^{2+}$ spins (S = 3/2), which is around 11.5 J/mol·K. The estimated value of $S_{mag}$, is considerably lower than the expected value of high spin Co$^{2+}$ spins (S = 3/2). It is worth noting that at 20 K, $S_{mag}$ = 5.6 J/mol·K, which closely matches the expected value (Rln2) of low spin state Co$^{2+}$ spins (S = 1/2) for the degrees of freedom of Ising-like moments. Therefore, the entropy associated with the Ising-like spins is primarily released below 20 K. The intrachain ferromagnetic (FM) interaction is much stronger than the interchain antiferromagnetic interaction in α-CoV$_2$O$_6$ and the intrachain spin degrees of freedom become effectively frozen, causing the spins to behave like Ising spin-1/2. This results in a significant reduction in magnetic entropy, and it indicates that magnetic fluctuations persist well above $T_N$. The missing magnetic entropy above the long-range order (LRO) state may be attributed to short-range magnetic ordering above $T_N$, which arises from the low-dimensional nature of this system [28]. The presence of SRO agrees with the result of dc magnetization studies. Indeed, such reduced entropy has been observed in other frustrated triangular lattice antiferromagnet systems [29].

To verify the crystal symmetry concerning T, temperature-dependent PXRD was performed in the temperature range of 10-300 K, and the PXRD pattern in a given temperature range is shown in the figure. S3 in the supplemental materials [25]. The Rietveld refinement was done of the PXRD pattern in each temperature range, and it was confirmed that no global structural phase transition in the given temperature range maintains the same monoclinic $C2/m$ crystalline symmetry. All the lattice parameters and cell volume decreased by lowering the temperature as shown in Figure. 4(a-c) and 4(d) respectively. A detailed analysis of the lattice parameters shows anomalies near $T_N$, a positive slope occurs, which indicates a positive thermal expansion. These elastic changes at the $T_N$ signify the existence of magnetoelastic coupling which correlates with the changes in the lattice dielectric constant. The thermal measurements at zero field show pronounced anomalies at $T_N$ in the thermal expansion coefficient and the relative length changes. These anomalies indicate significant spontaneous magnetostriction at $T_N$, highlighting the strong magnetoelastic coupling present in α-CoV$_2$O$_6$. This observation confirms the magnetoelastic coupling, which plays a vital role in the manifestation of the observed magnetoelectric coupling in this system. The α-CoV$_2$O$_6$ undergoes expansion upon heating, exhibiting positive thermal expansion. The temperature-dependent volume due to thermal expansion can be expressed as:

$$V(T) = V_0 \left[1 + \frac{A}{(e^{\left(\frac{\theta_D}{T}\right)}-1)}\right] \tag{3}$$

where, $V_0$ is cell volume extrapolated to 0 K, $\theta_D$ is the Debye temperature, and A is an adjustable fitting parameter. The Debye temperature obtained is $\theta_D \sim$ 446 K, which closely matches the value of the derived $\theta_D \sim$ 450 K from the specific heat results. Importantly, the temperature-dependent volume shows a small deviation below 80 K as shown in Figure. 4(d). A short-ranged magnetic ordering occurs below 100 K that changes the trend of lattice expansion owing to local symmetry changes without undergoing a structural phase transition [30].

Several low-dimensional magnetic systems were reported to exhibit interesting phenomena, which are related to the correlation between phonon and magnetic degrees of freedom. This motivated us to study the lattice dynamics of α-CoV$_2$O$_6$ by performing temperature-dependent Raman measurements over a wide temperature range of 5 - 180 K. In contrast to the diffraction techniques that examine the global structure of a crystal, Raman spectroscopy provides a distinctive method for detecting spin-phonon coupling, local structural phase changes, cationic ordering, etc.

According to group theory, monoclinic structure with space group C2/m possesses 12 Raman active modes with the irreducible representation $\Gamma_R = 8A_g + 4B_g$ [31,32]. Raman spectra of α-CoV$_2$O$_6$ were collected at various temperatures as shown in Figure. 5. Magnified views of different sections of the spectra are displayed in Figure. 6(a), where the Raman modes peak positions are denoted as P1-P13. Figure. 6(a). displays the Lorentzian function was used to fit the Raman spectra at each temperature and extract the phonon parameters. At the lowest temperature (T = 5 K), a total of 12 Raman modes are observed, as denoted by P1-P13. We also see a weak shoulder mode (P12) whose origin is not known at present. This mode is not expected as per group theory but may be due to disorder-induced Raman activity.

All the Raman spectra follow similar patterns in the temperature range of 5-180 K. The maximum modes belong to the $A_g$ modes. At lower frequencies, the two modes P1 and P2 (< 200cm$^{-1}$) originate from symmetric and antisymmetric stretching vibration, respectively, due to V-O bonds of edge-sharing between VO$_6$ octahedral pairs. Two modes P4, and P7 appearing at 262.3 cm$^{-1}$ and 344.2 cm$^{-1}$ are generated from CoO$_6$ and related to the lattice mode. The mode at P6 (304 cm$^{-1}$) originates from VO$_3$ mode (from the sharing of VO$_6$ octahedra between neighboring double chains). Two Raman modes P8 (432 cm$^{-1}$) and P11 (792.7 cm$^{-1}$) originate from the stretching vibration of the V-O-V bonds along the VO$_6$ octahedra parallel to the b-axis. In the 500-700 cm$^{-1}$ range, the Raman modes are attributed to the stretching vibrations of V-O bonds of two VO$_6$ in one chain. The most intense Raman mode at 886 cm$^{-1}$ corresponds to the stretching vibrations of the V-O bonds [33,34].

In general, the temperature-dependent Raman mode frequency change can be expressed as follows [35,36]:

$$\omega(T) = \omega_0 + \Delta\omega_{anh} + \Delta\omega_{SPC} + \Delta\omega_{EPC} \qquad (4)$$

where $\omega_0$ harmonic frequency of phonon modes that is obtained by extrapolating the experimental data down to 0 K, $\Delta\omega_{anh}$ indicates the pure anharmonic contribution from phonon-phonon interactions. Here $\Delta\omega_{SPC}$ and $\Delta\omega_{EPC}$ are the spin-phonon and electron-phonon coupling terms, respectively. The $\Delta\omega_{EPC}$ contribution is neglected here for the insulating nature of this material.

The anharmonic contribution ($\omega_{anh}$) is mainly due to phonon-phonon interactions, and changes in electronic structure with the temperature of higher-order phonon–phonon interactions. The shift in frequency due to an anharmonic contribution [37,38] is given by

$$(\omega)_{anh} = (\omega)_0 + A\left(1 + \frac{2}{e^{\frac{\hbar\omega_0}{2KT}}-1}\right) \qquad (5)$$

Here $\omega_0$ is the frequency of the phonon at absolute zero temperature, A is the cubic anharmonic coefficients for the frequency. $\hbar$ denotes the reduced Planck's constant and K is the Boltzmann constant, and T is the variable temperature. The right-hand second term corresponds to cubic anharmonicity, i.e., a three-phonon process that describes the decay of one optical phonon into two acoustic phonons of equal frequency.

Interestingly, for P2, P4, P5, P6, P7 and P10 a noticeable anomalous phonon softening is visible in the vicinity of $T_N$, which is shown in Figure. 6(b). The anomalous phonon softening below $T_N$ is due to the spin-phonon coupling effect or magnetostriction effect because of the absence of the role of electron-phonon coupling (insulating nature of its material) and no structural phase transition. A remarkable anomaly at $T_N$ in its thermal expansion measurement indicates the magnetostriction effect [4]. The variation of the phonon frequency with temperature for a few modes is shown in the Figure. 6(b). The behavior of most of the modes is captured by the normal cubic anharmonic equation (eqn. 5), however, for the modes labelled as P2, P4, P5, P6, P7 and P10 we do see a very weak deviation from the cubic anharmonic behavior below 100K, i.e., close to the temperature where short-range magnetic correlations begins in the system, thus indicating a possible presence of spin phonon coupling in α-CoV$_2$O$_6$. The magnetization and specific heat data corroborate the enhancement of SRO below 100 K.

As the phonons couple with the spin degrees of freedom, phonon renormalization occurs because of the modulation of the spin correlation <$S_i$ $S_j$> (which represents the statistical mechanical average for two neighboring spins $S_i$ and $S_j$). The frequency shift of phonon due to the SPC effect is given by: $(\Delta\omega)_{SPC}$ = - λ <$S_i$ $S_j$> = - λ $S^2$ φ(T) where φ(T) and λ represent the short-range order parameter and spin-phonon coupling (SPC) constant. According to the mean-field theory and the two-spin cluster method, Lockwood and Cottam determined the value of φ(T) for two AFM systems, such as FeF$_2$ (S = 2) and MnF$_2$ (S = 5/2) [39]. The value of φ(T) does not change with a change in the value of spin S, it was also used reasonably for S = 1 (NiF$_2$ and NiO) antiferromagnets [40,41]. Thus, for this material (S = 1/2) of Co$^{2+}$, the reported value of φ(T) can be used for the estimation of spin-phonon coupling (λ) using the following relation:

$$\lambda = -\left(\frac{(\omega(T)_{low} - \omega_{anh}(T)_{low})}{\varphi(T)_{low} - \varphi(2T)_N)S^2}\right) \quad (6)$$

where $\omega(T)_{low}$ and $\omega_{anh}(T)_{low}$ denote the experimental phonon frequency at the measured lowest temperature (in our case $(T)_{low}$ = 5 K) and the corresponding anharmonic estimate of the phonon frequency at the same temperature respectively. The obtained SPC values for the modes P2, P4, P5, P6, P7 and P10 using the equation. (6) are 0.28, 1.96, 3.43, 1.5, 2.8 and -1.92 cm$^{-1}$ respectively. The SPC value (λ) can be compared with some reported AFM compounds, such as MnF$_2$(0.4 cm$^{-1}$) [39], 15R – BaMnO3 (1.2-3.8 cm$^{-1}$) [38], Ni$_2$NbBO$_6$ (-0.3 to + 0.67 cm$^{-1}$) [42] and NiO (-7.9 cm$^{-1}$ and 14.1 cm$^{-1}$ for TO and LO phonons, respectively) [41], thus suggesting a reasonable value of SPC effect in α-CoV$_2$O$_6$ (-1.92 to 3.43). The fitting parameters of the phonon frequencies ($\omega_0$, A) and the spin-phonon coupling ($\lambda_{SPC}$) of the various Raman modes as shown in Table I.

## B. Theoretical Calculation for magnetoelectric coupling

Previous studies on α-CoV$_2$O$_6$ display magneto-dielectric coupling in the material, where a dielectric anomaly coincides with the onset of long-range magnetization at T$_N$ = 15 K [3]. In contrast, the temperature dependence of dielectric permittivity follows an opposite trend to that of magnetic susceptibility. While magnetic susceptibility decreases below T$_N$, the dielectric permittivity increases instead of decreasing. This behavior demonstrates the coupling between electrical charges and magnetic ordering. These observations suggest the possibility of magnetoelectric coupling, which could arise from one or more mechanisms, including exchange-striction, charge transfer and spin-driven interactions. To investigate the impact of magnetism on the electronic properties, we computed the charge density difference between the AFM and NM states, defined as $\Delta\rho = \rho_{AFM} - \rho_{NM}$. In Figure. 7(b) shows that the majority of the charge density is concentrated around the CoO$_6$ octahedra. Oxygen atoms can be distinguished based on their positions within the unit cell: (1) those bonded within the CoO$_6$ octahedra and (2) those not bonded to the CoO$_6$ octahedra. Quantitatively, the average charge difference between the AFM and NM is 0.0403e for oxygen atoms bonded within the octahedra, for non-bonded oxygen atoms, and 0.135e for Co atoms, which lose charge. Furthermore, Table II. provides the charge transfer values for

Co and O atoms within the $CoO_6$ octahedra in the unit cell. This indicates that the electric charge transfer occurs between the oxygen and Co atoms within the $CoO_6$ octahedra, while the other oxygen ions contribute negligibly to this charge transfer as shown in Table III. This suggests that Co-O coupling and the $O^{2-}$ atoms in the octahedra contribute to the development of the observed long-range magnetic order. A similar charge-transfer-driven multiferroic effect has been observed in the multiferroic system [43]. Additionally, we note that there is almost no charge density difference around the V atoms. Furthermore, to support the involvement of the O atoms, we present a detailed orbital analysis in Figure. 7(b). The charge density difference in Figure. 7(b) highlights the importance of Co-O interactions, prompting us to further investigate the orbital-projected density of states. The density of states of individual atoms is plotted as shown in Figure. 7(c) and 7(d) for the AFM states by considering without and with SOC effect respectively. The spins in the magnetic structures of this material are arranged in a collinear configuration as shown in Figure. 7(a). The total magnetic moment of 0 $\mu_B$ per atom is obtained for the AFM state with the Co atoms having a magnetic moment of ~$2.7\mu_B$, consistent with previous studies [12,17]. The O - p orbitals and Co – d orbitals have significant contributions near the Fermi level while the contributions of V are minimal in both cases, as shown in Figures. 7(c) and 7(d). The O-p and Co-d states overlap in both cases, indicating strong O-p and Co-d hybridizations. This observed p-d hybridization is recognized as one of the spin-driven mechanisms that can give rise to multiferroic properties. But due to the centrosymmetric space group of α-$CoV_2O_6$, it didn't show any measurable electric polarization [3], so there is no possibility of charge-transfer-induced ferroelectricity. As shown in Figure. 7(b), the charge transfer between the Co and O atoms in the $CoO_6$ octahedra is symmetric, which may result in the cancellation of the total electric dipole moment across the entire unit cell, leading to net-zero electric polarization which validates the reported experimental result. We have obtained a higher value of charge in the AFM state than the NM state. This evidence proves the assumptions that the dielectric permittivity increases below $T_N$, which couples the magnetic ordering to the electrical charges. The combined evidence from magnetic ordering and lattice parameter anomalies suggests that magnetoelastic phenomena play a crucial role in effectively coupling the magnetic spins with the electrical dipoles at $T_N$. The DFT results for the AFM ground state below $T_N$ shows a strong hybridization between the Co-d and O-p states, in contrast to the NM state, which supports the role of p-d hybridization in the observed dielectric anomaly. Moreover, a clear charge transfer between the Co and O atoms indicates that the redistribution of local charges due to magnetic ordering affects the lattice parameters leading to the dielectric features observed at $T_N$. This study emphasizes the intricate relationship between magnetism and electrical properties in α-$CoV_2O_6$.

## CONCLUSION

The quasi-one-dimensional Ising spin chain system α-$CoV_2O_6$ has been extensively studied by DC magnetization, specific heat, temperature-dependent PXRD and temperature-dependent Raman spectroscopy measurements. In contrast to most low-dimensional spin-chain compounds, α-$CoV_2O_6$ demonstrates long-range antiferromagnetic ordering at a specific temperature, i.e. $T_N$ = 15 K. From specific heat data, missing entropy indicates the onset of short-range magnetic ordering well above the $T_N$ was observed. The temperature-dependent XRD shows the variation of the lattice parameters concerning temperature, indicating the magneto-elastic coupling below $T_N$. We observe phonon anomalies appear due to short-range magnetic ordering (below 100 K) well above the $T_N$ (= 15 K), thus signifying the presence of spin-phonon coupling. The ab initio DFT calculations predicted the ground state as insulating AFM of the system. Moreover, the interaction of spin and charge through the magneto-dielectric effect below $T_N$ is confirmed by our theoretical calculation, which indicates that strong p-d hybridization facilitates charge transfer between the Co-d and O-p orbitals. The charge redistribution associated with magnetic ordering and structural anomaly points to a complex interplay of spin, phonon degrees of freedom, and charge in the quasi-one-dimensional spin-chain system of α-$CoV_2O_6$.

## ACKNOWLEDGMENTS

D.N. acknowledges IISER Kolkata for senior research fellowship (SRF) and the instrumental facilities.

**Figure 1**

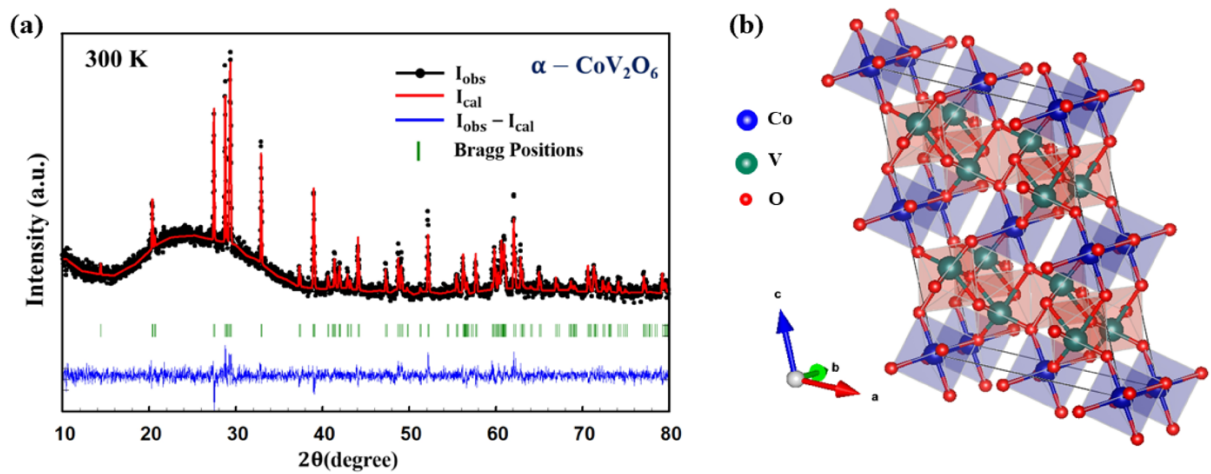

FIG. 1. (a) Rietveld refinement of room-temperature x-ray diffraction of powdered α-CoV$_2$O$_6$; (b) the crystal structure of α-CoV$_2$O$_6$ in which blue octahedra denote the Co$^{2+}$ in the b-axis direction with CoO$_6$ atomic arrangement forming a chain and V$^{5+}$ is shown in green color and oxygen in red color.

**Figure 2**

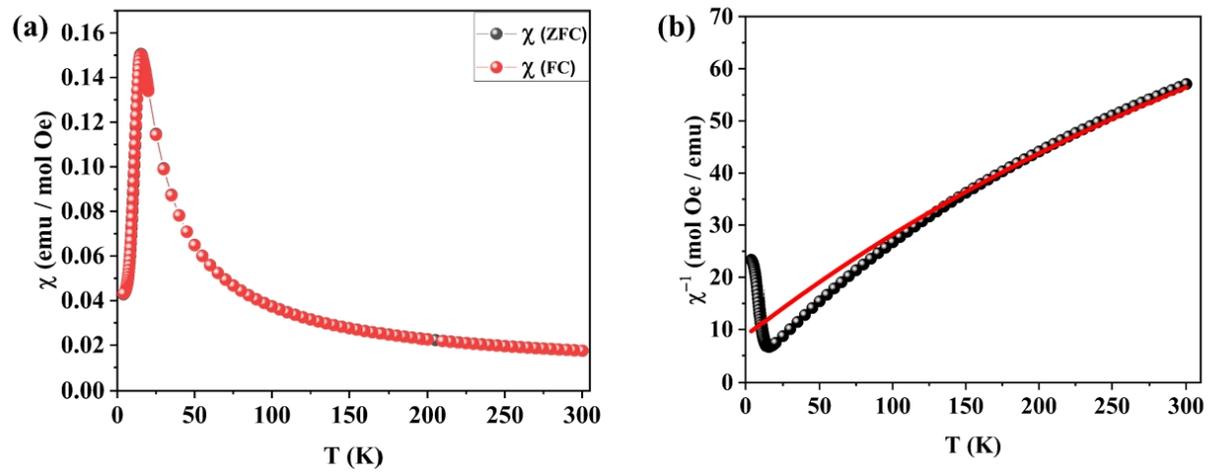

FIG. 2. (a) ZFC-FC susceptibility and (b) inverse susceptibility of α-CoV$_2$O$_6$ at H = 1 T.

**Figure 3**

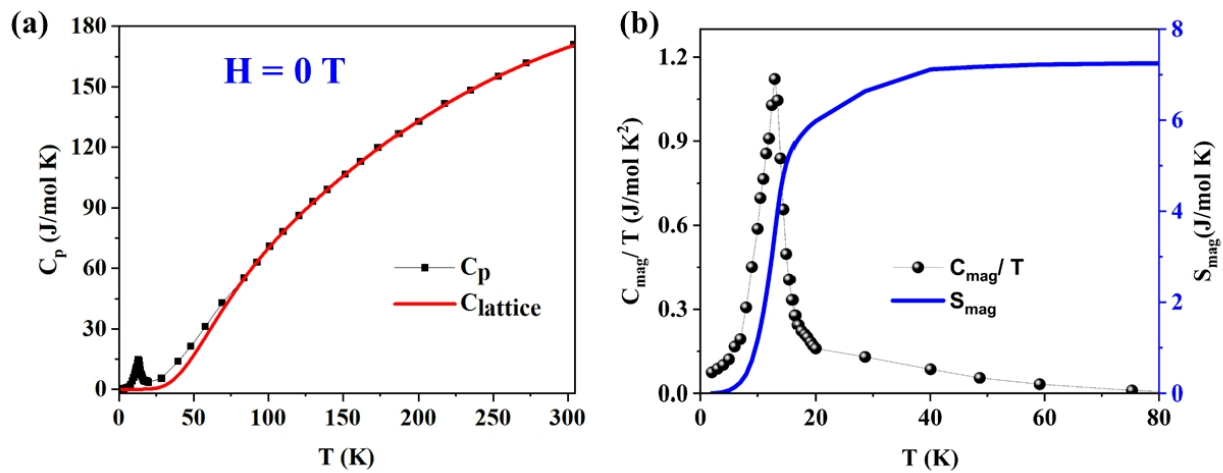

FIG. 3. (a) Specific heat curve measured at zero magnetic fields. The red line represents the phononic ($C_{lattice}$) of $C_P$ (T) ;(b) $C_{mag}$ / $T$ curve as a function of temperature and its corresponding magnetic entropy $S_{mag}$.

**Figure 4**

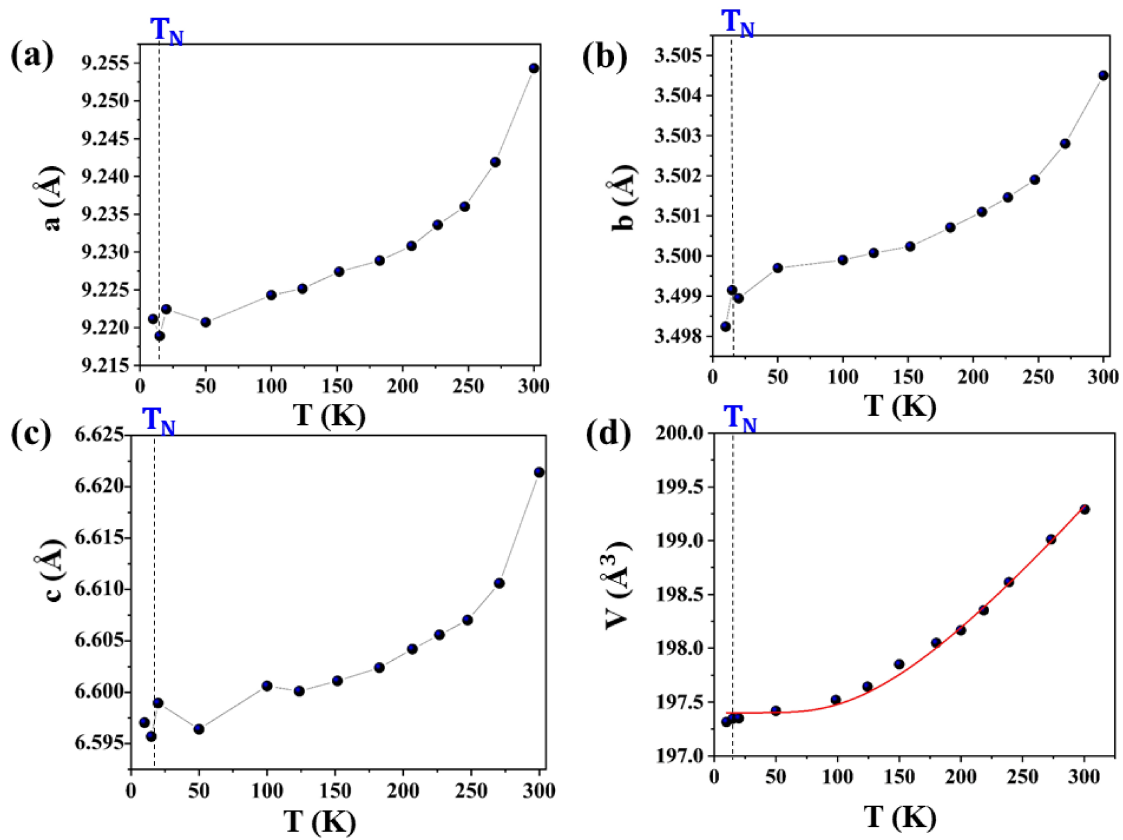

FIG 4. (a-c) Temperature variation of the lattice parameters and (d) volume (V); dashed lines indicate changes in lattice parameters near $T_N$.

**Figure 5**

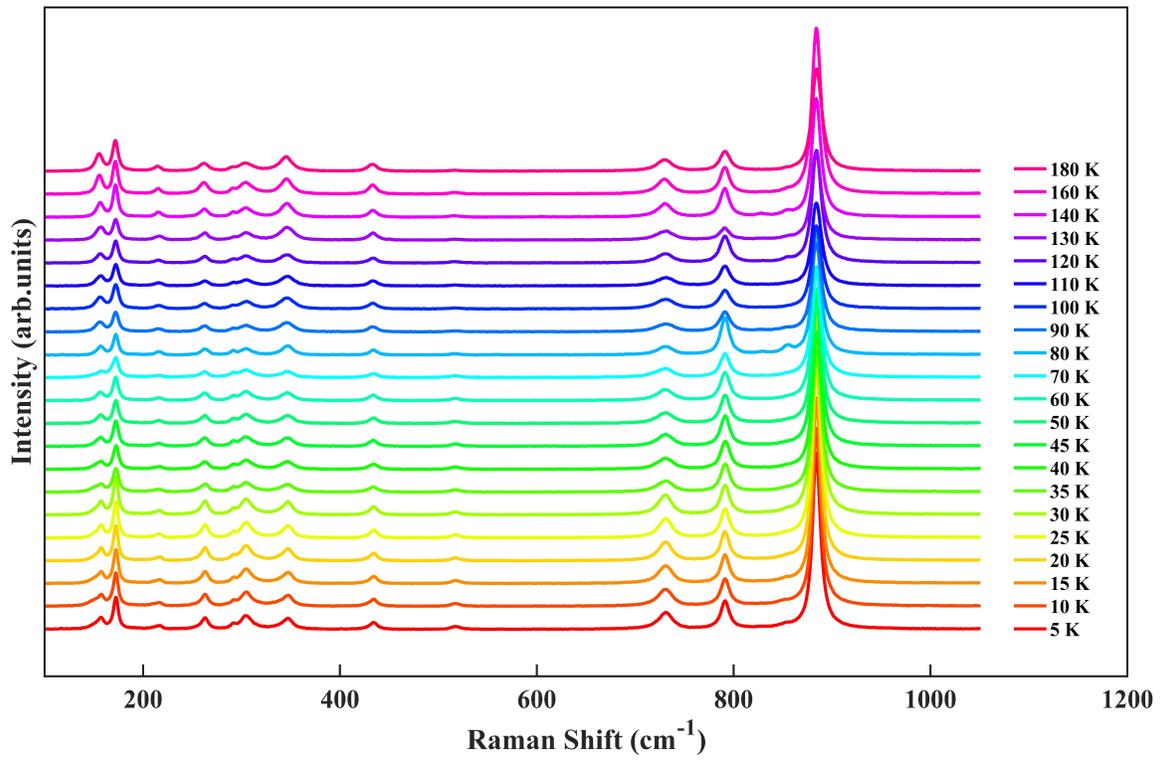

FIG. 5. Raman spectrum of α-CoV$_2$O$_6$ at different temperatures.

**Figure 6**

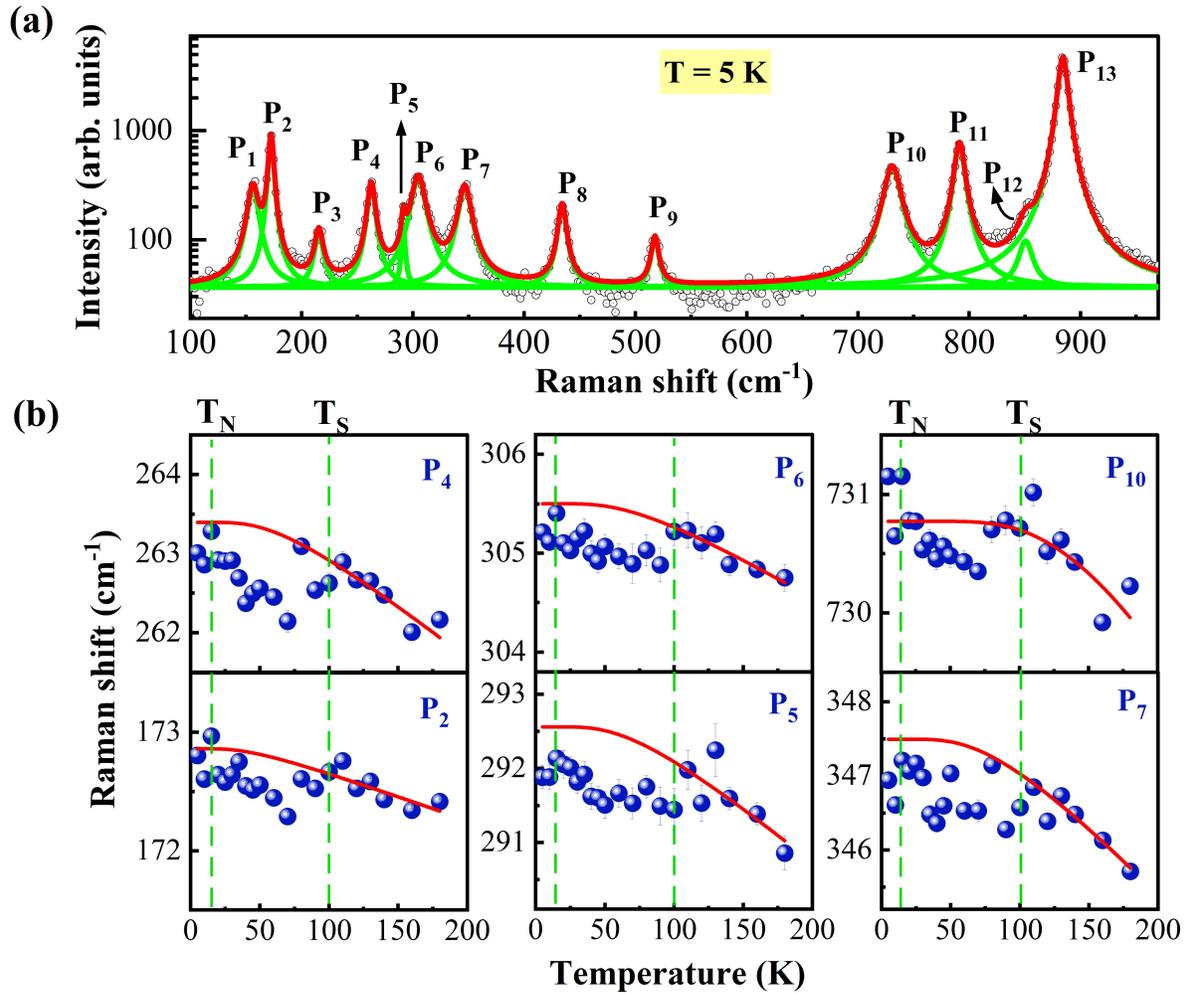

FIG. 6. (a) Lorentzian fitting of the Raman spectrum of α-CoV$_2$O$_6$ collected at 5 K, (b) Thermal variation of Raman shifts for the P2, P4, P5, P6, P7, and P10 modes and clear deviation below $T_S$ and $T_N$. The red line is fitted with equation (5) showing deviation in anharmonic behavior below short-range magnetic ordering well above the $T_N$.

**Figure 7**

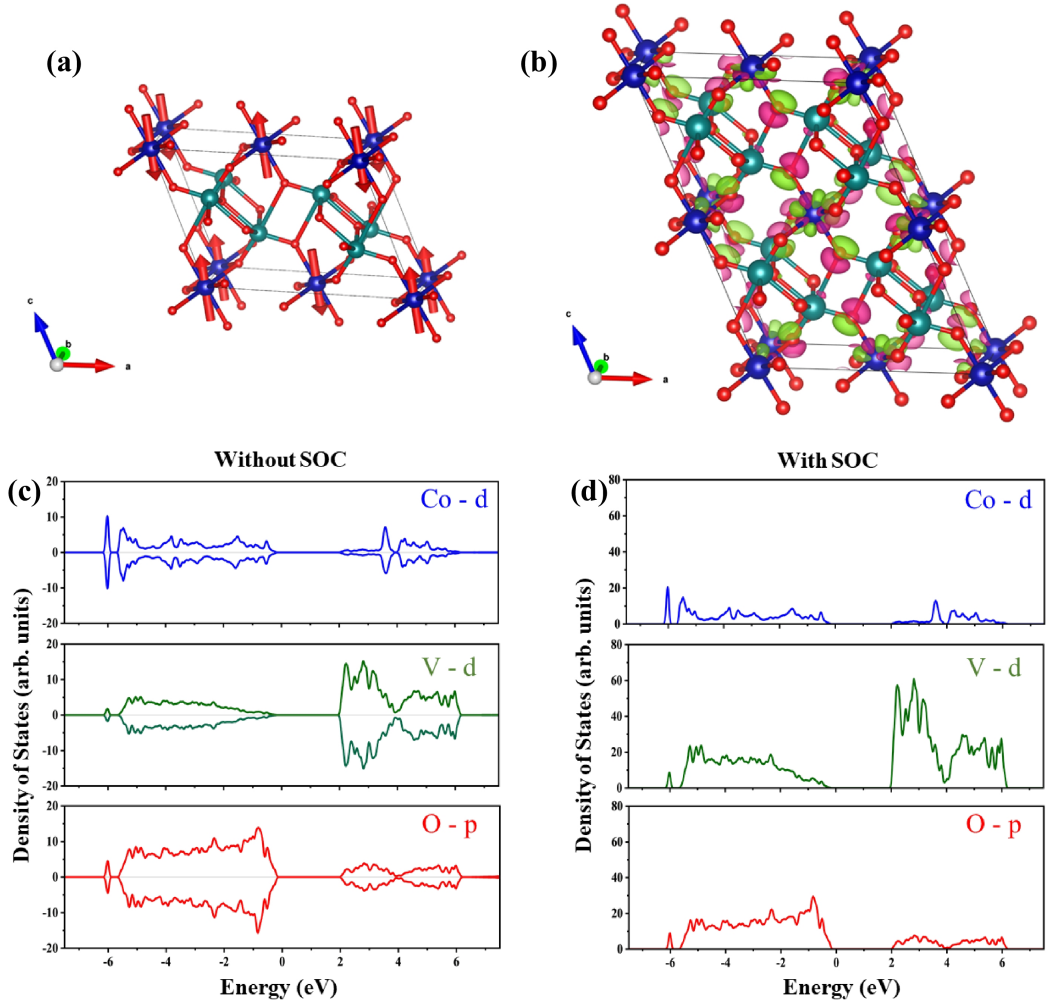

FIG. 7. (a) The crystal structure of monoclinic α-CoV$_2$O$_6$ shown with arrows (spin-up and spin-down are indicated by red arrows) indicating the ordered magnetic directions at the Co$^{2+}$ ions in AFM state. Co, V are shown for the blue and green spheres and O are small sphere with red color; (b) charge density difference of α-CoV$_2$O$_6$ and visualization of the charge distribution by using VESTA software; The density of states of the AFM α-CoV$_2$O$_6$ (c) with-out and (d) with SOC.

**TABLE I.** The fitting parameters of the phonon frequencies ($\omega_0$, $A$) and the spin-phonon coupling ($\lambda_{SPC}$) of the various Raman modes.

| Modes (cm$^{-1}$) | $\omega_0$ (cm$^{-1}$) | $A$ (cm$^{-1}$) | $\lambda_{SPC}$ |
|---|---|---|---|
| P1 | 173.12 ± 0.28 | -0.26 ± 0.11 | 0.28 |
| P2 | 264.76 ± 0.65 | -1.36 ± 0.37 | 1.96 |
| P3 | 294.28 ± 0.87 | -1.72 ± 0.54 | 3.43 |
| P4 | 306.46 ± 0.38 | -0.96 ± 0.24 | 1.5 |
| P7 | 350.15 ± 0.56 | -2.65 ± 0.40 | 2.8 |
| P10 | 738.13 ± 3.95 | -7.36 ± 3.74 | -1.92 |

TABLE II. The quantity of Badger charge transfer was calculated via density functional theory for both the AFM and NM states for the Co and O atoms with charge differences.

| Atoms | Charge (e) (AFM case) | Charge (e) (NM case) | Charge (e) (=AFM-NM case) |
|---|---|---|---|
| Co1 | 1.400924 | 1.265499 | 0.135425 |
| O1 | -0.852104 | -0.802392 | -0.049712 |
| O2 | -0.852104 | -0.802392 | -0.049712 |
| O3 | -0.852071 | -0.802341 | -0.049730 |
| O4 | -0.852071 | -0.802341 | -0.049730 |
| O5 | -0.847749 | -0.826706 | -0.021043 |
| O6 | -0.847749 | -0.825838 | -0.021911 |
| Co2 | 1.400106 | 1.265499 | 0.134607 |
| O1 | -0.852422 | -0.802392 | -0.050030 |
| O2 | -0.852422 | -0.802392 | -0.050030 |
| O3 | -0.852389 | -0.802341 | -0.050048 |
| O4 | -0.852389 | -0.802341 | -0.050048 |
| O5 | -0.846145 | -0.826706 | -0.019439 |
| O6 | -0.846145 | -0.825838 | -0.020307 |
| Co3 | 1.400106 | 1.265499 | 0.134607 |
| O1 | -0.852389 | -0.802360 | -0.050029 |
| O2 | -0.852389 | -0.802360 | -0.050029 |
| O3 | -0.852422 | -0.802373 | -0.050049 |
| O4 | -0.852422 | -0.802373 | -0.050049 |
| O5 | -0.846145 | -0.825838 | -0.020307 |
| O6 | -0.846145 | -0.826706 | -0.019439 |
| Co4 | 1.400924 | 1.265499 | 0.135425 |
| O1 | -0.852071 | -0.802360 | -0.049711 |
| O2 | -0.852071 | -0.802360 | -0.049711 |
| O3 | -0.852104 | -0.802373 | -0.049731 |
| O4 | -0.852104 | -0.802373 | -0.049731 |
| O5 | -0.847749 | -0.825838 | -0.021911 |
| O6 | -0.847749 | -0.826706 | -0.021043 |

**TABLE III.** The quantity of Badger charge transfer calculated via density functional theory for both the AFM and NM states for the V and O atoms with the charge difference whose contribution is less in charge transfer.

| Atoms | Charge (e) (AFM case) | Charge (e) (NM case) | Charge (e) (=AFM-NM case) |
|---|---|---|---|
| V1 | 1.928190 | 1.925341 | 0.002849 |
| V2 | 1.927821 | 1.925341 | 0.002480 |
| V3 | 1.927821 | 1.925341 | 0.002480 |
| V4 | 1.928190 | 1.925341 | 0.002849 |
| V5 | 1.927821 | 1.925341 | 0.002480 |
| V6 | 1.928190 | 1.925341 | 0.002849 |
| V7 | 1.928190 | 1.925341 | 0.002849 |
| V8 | 1.927821 | 1.925341 | 0.002480 |
| O1 | -0.929062 | -0.929766 | 0.000704 |
| O2 | -0.929484 | -0.929766 | 0.000282 |
| O3 | -0.929076 | -0.929136 | 0.000060 |
| O4 | -0.928654 | -0.929136 | 0.000482 |
| O5 | -0.929484 | -0.929774 | 0.000290 |
| O6 | -0.929062 | -0.929774 | 0.000712 |
| O7 | -0.928654 | -0.929128 | 0.000474 |
| O8 | -0.929076 | -0.929128 | 0.000052 |

# Evidence of spin-phonon-charge coupling in quasi-one-dimensional Ising spin chain system α – $CoV_2O_6$

Debismita Naik,[1,*] Souvick Chakraborty,[1] Akriti Singh,[2] Ayan Mondal,[3] Surajit Saha,[2] Venkataramanan Mahalingam,[3] Satyabrata Raj,[1,4] and Pradip Khatua[1,†]

[1]Department of Physical Sciences, Indian Institute of Science Education and Research Kolkata, Mohanpur, Nadia, West Bengal 741246, India.

[2]Department of Physics, Indian Institute of Science Education and Research, Bhopal 462066, India.

[3]Department of Chemical Sciences, Indian Institute of Science Education and Research Kolkata, Mohanpur, Nadia, West Bengal 741246, India.

[4]National Centre for High-Pressure Studies, Indian Institute of Science Education and Research Kolkata, Mohanpur, Nadia, 741246, India

Corresponding authors: [*]dn17ip012@iiserkol.ac.in

[†]pradip.k@iiserkol.ac.in

# Room Temperature X-ray diffraction (XRD) Study

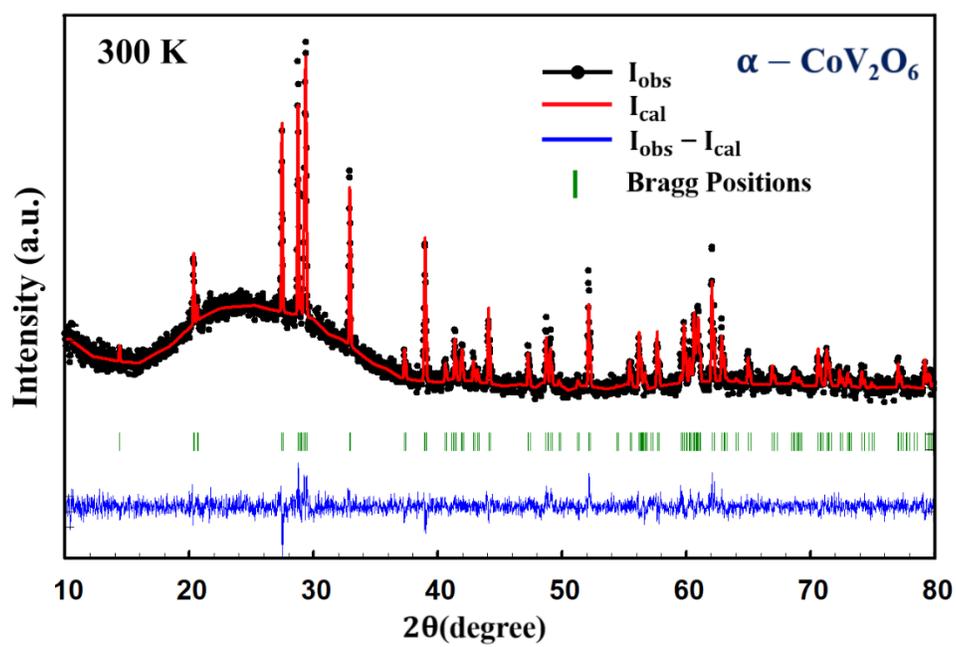

**Figure S1.** Rietveld refinement of XRD pattern at 300 K of α – $CoV_2O_6$.

**TABLE S1.** Structural parameters and atomic sites as determined from Rietveld refinement of the XRD data at 300 K and refined parameter $\chi^2 = 1.08$.

Space group: C2/m

| Sample Name | α – $CoV_2O_6$ |
| --- | --- |
| a (Å) | 9.2543 Å |
| b (Å) | 3.5045 Å |
| c (Å) | 6.6213 Å |
| V (Å³) | 199.454 Å³ |
| Atom | |
| Co | |
| x | 0 |
| y | 0 |
| z | 0 |
| V | |
| x | 0.3056 |
| y | 0.5 |
| z | 0.3378 |
| O1 | |
| x | 0.1535 |
| y | 0.5 |
| z | 0.1132 |
| O2 | |
| x | 0.4642 |
| y | 0.5 |
| z | 0.2742 |
| O3 | |
| x | 0.1915 |
| y | 0.5 |
| z | 0.5623 |

# Scanning electron microscopy (SEM) and elemental mapping

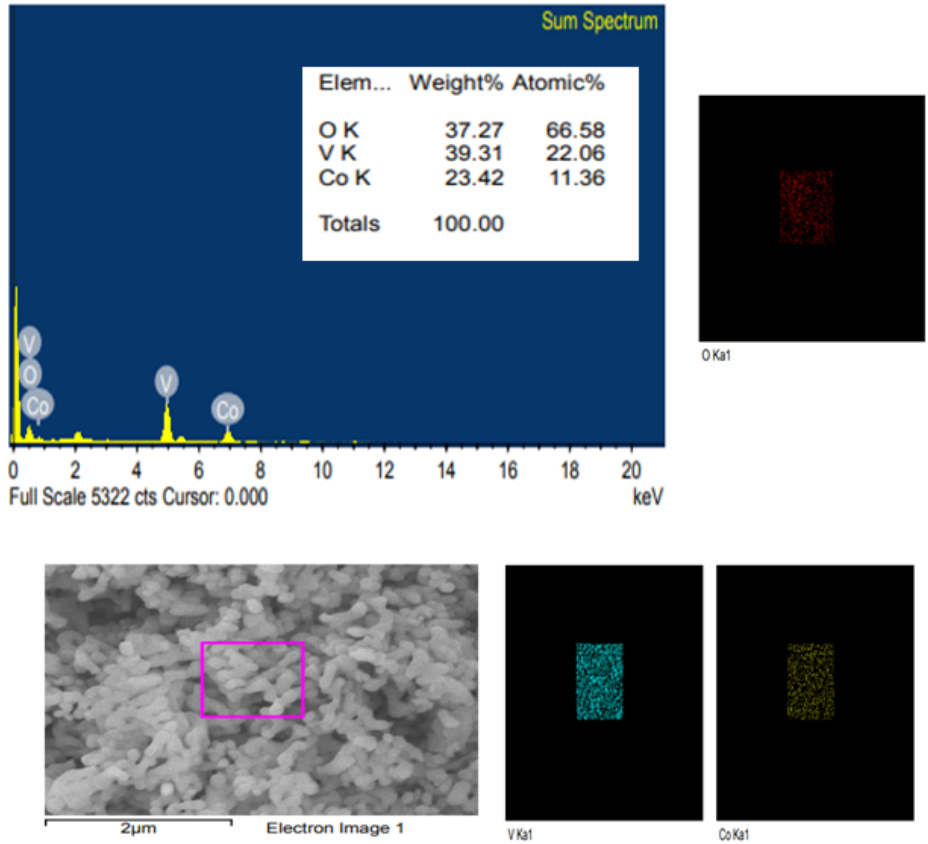

**Figure. S2.** SEM and elemental composition and mapping using Energy dispersive X-ray analysis (EDX).

## Temperature dependent Powder XRD (PXRD)

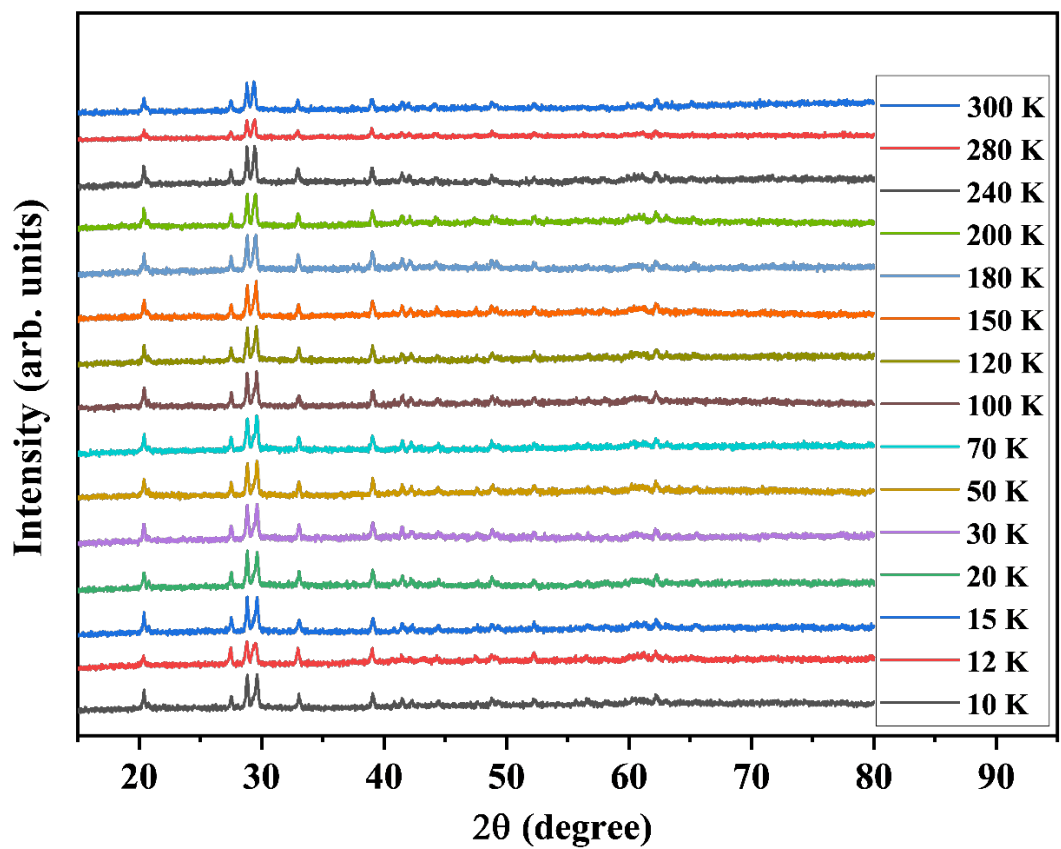

**Figure. S3.** Temperature-dependent PXRD pattern at different temperatures.